# Lensless multicore-fiber microendoscope for real-time tailored light field generation with phase encoder neural network (CoreNet)

JIAWEI SUN,[1,2,6,†] JIACHEN WU,[1,3,7,†] NEKTARIOS KOUKOURAKIS,[1,2] ROBERT KUSCHMIERZ,[1,2] LIANGCAI CAO[3] AND JÜRGEN CZARSKE, [1,2,4,5,8]

[1]*Laboratory of Measurement and Sensor System Technique, TU Dresden, Helmholtzstrasse 18, 01069 Dresden, Germany*
[2]*Competence Center for Biomedical Computational Laser Systems (BIOLAS), TU Dresden, Dresden, Germany*
[3]*State Key Laboratory of Precision Measurement Technology and Instruments, Department of Precision Instruments, Tsinghua University, Beijing 100084, China*
[4]*Cluster of Excellence Physics of Life, TU Dresden, Dresden, Germany*
[5]*Institute of Applied Physics, TU Dresden, Dresden, Germany*
[6]*jiawei.sun@tu-dresden.de*
[7]*wjc18@mails.tsinghua.edu.cn*
[8]*juergen.czarske@tu-dresden.de*
*† These authors contributed equally to this work*

**The generation of tailored light with multi-core fiber (MCF) lensless microendoscopes is widely used in biomedicine. However, the computer-generated holograms (CGHs) used for such applications are typically generated by iterative algorithms, which demand high computation effort, limiting advanced applications like in vivo optogenetic stimulation and fiber-optic cell manipulation. The random and discrete distribution of the fiber cores induces strong spatial aliasing to the CGHs, hence, an approach that can rapidly generate tailored CGHs for MCFs is highly demanded. We demonstrate a novel phase encoder deep neural network (CoreNet), which can generate accurate tailored CGHs for MCFs at a near video-rate. Simulations show that CoreNet can speed up the computation time by two magnitudes and increase the fidelity of the generated light field compared to the conventional CGH techniques. For the first time, real-time generated tailored CGHs are on-the-fly loaded to the phase-only SLM for dynamic light fields generation through the MCF microendoscope in experiments. This paves the avenue for real-time cell rotation and several further applications that require real-time high-fidelity light delivery in biomedicine.**

## 1. INTRODUCTION

A phased array, usually a computer-controlled array of transmitters, can control the orientation or shape of the output beam of waves, having wide applications in astronomy [1], radar technology [2], space communication [3], quantum communication [4], ultrasound technique [5], and optical engineering [6]. As an attractive tool for optical phased arrays, spatial light modulators (SLMs) facilitate holographic displays with high resolution. Hence, computer-controlled precise optical wavefront shaping is realized with SLMs, enabling new applications in microscopy [7], holographic optogenetic stimulation [8], optical manipulations [9], and imaging through biological tissues [10–12]. To break through the limit of the penetration depth of free beams [13], endoscopes become an essential tool for imaging hard-to-reach areas.

In clinical diagnostics, endoscopes are widely implemented for in-vivo imaging with low invasiveness and offer diameters of a few millimeters. Recently, MCFs are employed for lensless microendoscopic imaging for minimum invasiveness of a few hundred microns [14–16]. Each MCF contains thousands of single-mode cores, and each fiber core can function both as an imaging pixel for detection of light [14,17] and a light-emitting unit in a phased array for wavefront shaping [18]. When imaging through the MCF-based microendoscope, the discrete distribution of cores and the core-to-core spacing of the MCF limits the image resolution. Image reconstruction from speckle correlations is proved as an efficient method to further improve the imaging resolution [19,20]. The limited field of view was a trade-off for the needle size of lensless microendoscopes until recent advances on computational image recovery have demonstrated high-resolution widefield imaging through the microendoscope [16,21]. On the other hand, employing the lensless microendoscope as a remote phased array can generate a diffraction-limited focus in the far-field of the distal application side of the microendoscope by displaying a Fresnel lens on the proximal facet [18]. This is implemented for two-photon endoscopic imaging, which is realized by scanning the focus using the phase-only SLM [22].

Implementing an adaptive lens and Galvo mirror for rapid 3D-scanning can break through the low refreshing rate of the SLM, enabling video-rate 3D fluorescent imaging through the microendoscope [23]. The Gerchberg-Saxton (GS) algorithm [24] is commonly utilized for calculating the CGH for structured light field generation. However, the random and discrete distribution of fiber cores leads to spatial sampling of the CGH when coupling into the MCF, and the number of fiber cores is much less than the pixel number of the phase-only SLM, inducing strong aliasing. This results in unrecognizable reconstructed light fields [25], proving the GS algorithm cannot be directly implemented in hologram generation for randomly distributed phased-arrays. Therefore, we previously proposed a tailored phase retrieval algorithm for holographic control of complex light field through the MCF named Core-GS [25], enabling complex wavefront shaping through the MCF with high fidelity. This is implemented in lab-on-a-chip optical manipulation of biological cells [26] and can facilitate applications in holographic optogenetic stimulation [27], micro-materials processing in hard-to-reach areas [28], structured light generation for MCF amplifier [29], and high-dimensional optical and quantum communication [30]. Due to the long computation time, the CGHs have to be generated in advance and then loaded to the SLM for dynamic light field generation. This limits the application such as adaptive tomographic optical manipulation [9,31] and adaptive selective holographic photoactivation in optogenetics [32], which needs rapid generation and refreshing of CGHs. Therefore, there is a great demand in biomedicine for real-time generation of tailored CGHs for MCF. Recently, deep neural networks have been used for reducing the computation time of CGHs [33–37], however, all of these existing networks are designed for regularly distributed pixels and induces significant distortion when applied to discrete or randomly distributed phased-arrays [25]. Therefore, an approach that can generate tailored CGHs for MCFs or random phased arrays in real-time with high fidelity is not yet available.

In this paper, we propose a novel phase encoder neural network (CoreNet), which can generate CGHs tailored for MCFs in real-time. The discrete or random distribution map of the phased array can be loaded to the neural network as an input. Hence, CoreNet can generate tailored phase modulation holograms rapidly, enabling precise light-field control for a randomly distributed phased array. Specifically, for holographic display through a lensless microendoscope, the experimentally measured fiber core distribution map is embedded into the neural network to generate tailored CGHs. Unlike supervised learning, which needs to label the holograms calculated by classical phase retrieval algorithms as the ground truth, CoreNet uses an autoencoder network architecture to encode the target intensity into the tailored phase modulation holograms with unsupervised training. The diffraction model is incorporated into the network for numerically propagating the light field between the phase modulation plane and the target intensity plane. This allows the network to search for the optimal phase modulation maps of the target image and learn the mapping from the target images to the phase modulation holograms. This end-to-end architecture of CoreNet gets rid of complicated iterative operations, and can rapidly generate the phased array after the training. Near video-rate generation of the tailored CGHs for MCF-based complex wavefront shaping can thus be realized by the trained network with high fidelity, opening new perspectives for applications based on MCFs.

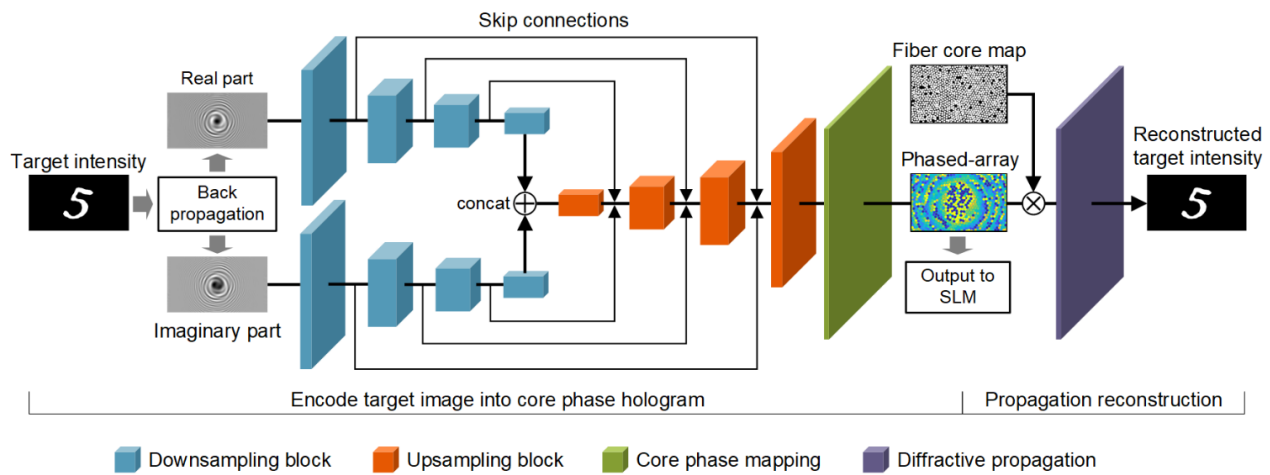

**Fig. 1.** Structure of the phase encoder deep neural network (CoreNet). The encoding part of CoreNet is a modified U-Net. The downsampling path of U-Net is split into two paths, and the inputs of the two paths are the real and imaginary parts of the field at the distal facet of the fiber bundle, which is obtained by back-propagation of the target field to the plane of the distal fiber facet. The core phase mapping and diffractive propagation are embedded in CoreNet to reconstruct the target field. Since the labeled data is the same as the input target field, CoreNet can achieve unsupervised learning.

## 2. RESULTS

### A. Principle of CoreNet

When light field transmits through the MCF, the intrinsic optical path differences (OPDs) between fiber cores induce strong phase distortion in the light field coupled out. To compensate for this, digital optical phase conjugation (DOPC) [38,39] is employed. The phase differences between fiber cores are measured by off-axis digital holography [26], and the conjugated phase differences are mapped on the MCF facet by the phase-only SLM to pre-compensate the OPDs. Then the MCF can act as a phased array to generate arbitrary light field distribution. Typically, to control the wavefront through the MCF, a tailored CGH generated by the Core-GS algorithm [25] is loaded to the SLM additionally to the conjugated phase. Although the Core-GS algorithm can achieve good quality compared to the raw GS algorithm, the long iteration time still limits the applications.

We designed a phase encoder neural network, called CoreNet, to generate tailored holograms for high-speed complex wavefront shaping through MCF. The U-Net architecture has been proven to be effective in image processing tasks. Here, we modified the U-Net to a network with two inputs to collect more information on the target image (Fig. 1). Instead of feeding the raw image directly to the neural network, the target image firstly back propagates to the plane on the fiber facet at the distal application side to obtain the complex amplitude. The reason for that is learning the feature mapping at the same plane is easier than at different planes. Then the real and imaginary parts of the complex amplitude are extracted as the inputs of CoreNet. The downsampling blocks are duplicated for each input. The two downsampling paths join a bottleneck layer by a concatenation operator. Then the bottleneck layer is up-sampled to the resolution of the phase-only SLM by a series of upsampling blocks.

Each downsampling and upsampling block consists of two residual blocks as shown in Fig. 2. Each residual block is composed of two sets of batch normalization (BN), nonlinearity (ReLU), and a convolutional layer stacked one above the other. The strides of the first convolution and transposed convolution in the residual block are (2, 2) to realize the function of downsampling and upsampling.

At the end of U-Net, the values at the core position are extracted to convolve with circular masks which indicate the shape of fiber cores to form the tailored phase modulation map for the phased array. Then the fiber core map as amplitude combined with phased array pattern to form a complex field. Finally, the complex field propagates to the target plane to form the target intensity distribution. Here we adopted band-limited angular spectrum method [40] to simulate the light propagation. Comparing with the classical angular spectrum method, it has less numerical error for far-field propagation. The band-limited angular spectral transfer function is

$$U_o(x_o, y_o) = F^{-1}\{H_{\text{limit}}(u,v) \cdot F\{U_i(x_i, y_i)\}\} \quad (1)$$

$$H_{\text{limit}}(u,v) = \exp\left(i2\pi z\sqrt{\frac{1}{\lambda^2} - u^2 - v^2}\right)\text{rect}\left(\frac{u}{2u_{\text{limit}}}, \frac{v}{2v_{\text{limit}}}\right) \quad (2)$$

$$u_{\text{limit}} = \left[(2\Delta u z)^2 + 1\right]^{-1/2} \lambda^{-1}$$
$$v_{\text{limit}} = \left[(2\Delta v z)^2 + 1\right]^{-1/2} \lambda^{-1} \quad (3)$$

where $\Delta u$ and $\Delta v$ denote the sample interval in the frequency domain.

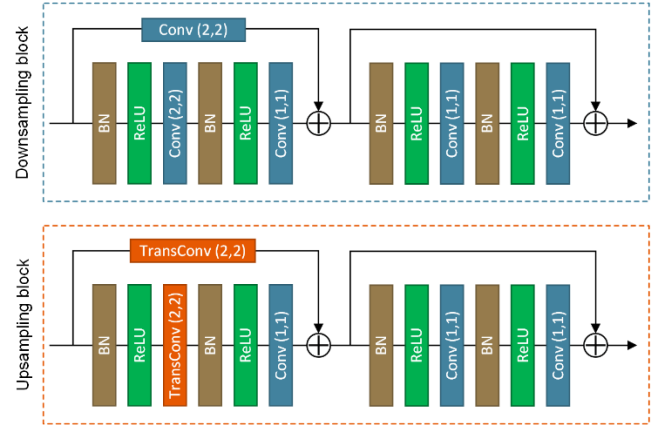

**Fig. 2.** The principle of the downsampling and upsampling block. The numbers in parentheses indicate the strides of convolution layers and transposed convolution layers.

As an unsupervised learning method, the corresponding phased array of each input image does not need to be calculated in advance. By utilizing automatic differentiation, the loss can be propagated back to the encoder part, and the learnable parameters of the U-net can be updated during the training process. The training results are shown in Fig. 3. We use the EMNIST Balanced dataset which includes 131,600 characters for training [41]. Several custom letters and binary patterns make up the test set. 16 images and 2 images are randomly selected from the MNIST validation dataset and custom test dataset to show the network performance. The reconstructed results by the Core-GS algorithm are also presented for comparison. CoreNet can achieve better reconstruction quality and the computation time is only 0.2 s, which is much faster than the Core-GS algorithm that took 11s on the same platform (see "Materials and methods".). The 2-D correlation coefficient is employed to characterize the fidelity of reconstructed images. Hence, for a normalized image $X$, the correlation coefficient between the reference image $Y$ is expressed as

$$\text{cc} = \frac{\sum_{i}^{n}(X_i - \bar{X})(Y_i - \bar{Y})}{\left\{\sum_{i}^{n}(X_i - \bar{X})^2 \sum_{i}^{n}(Y_i - \bar{Y})^2\right\}^{1/2}}, \quad (4)$$

where $\bar{X}$ and $\bar{Y}$ is the mean value of the reconstructed image and the reference image, $n$ is the total number of pixels.

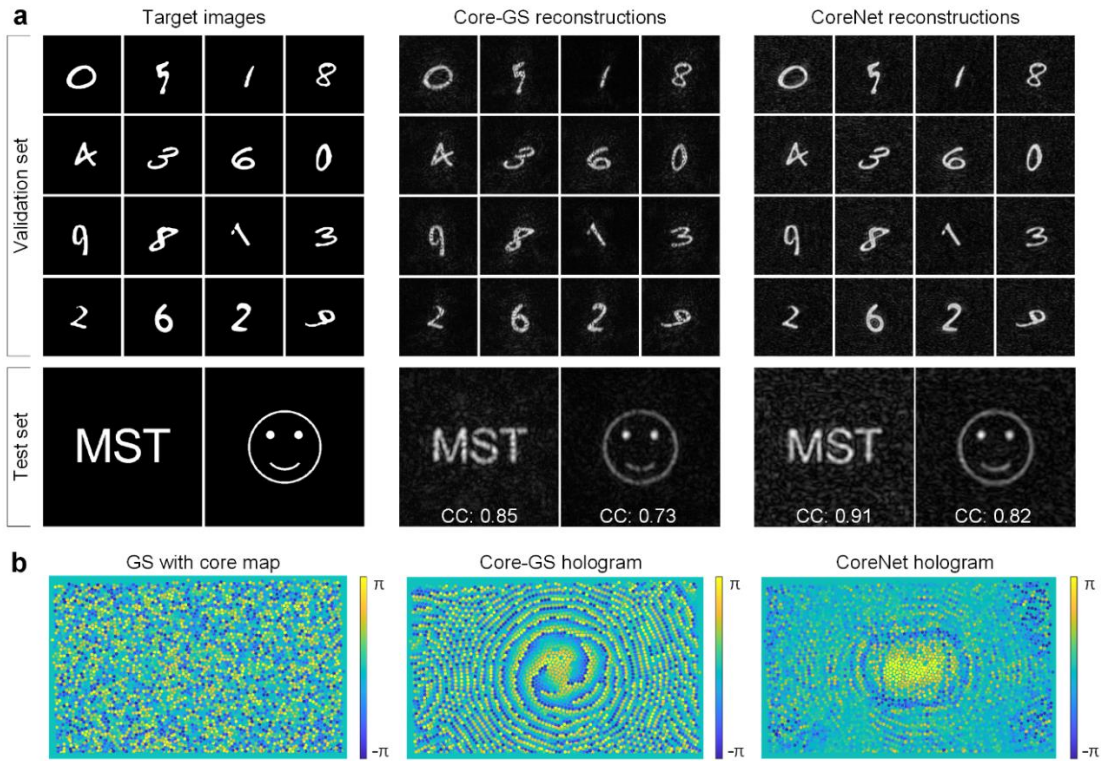

**Fig. 3.** Training and validation data for CoreNet. (a) The EMNIST dataset is employed to train CoreNet. The custom dataset containing letters and patterns is employed to test CoreNet, and the letter "MST" and a smiling face are shown to demonstrate the performance of the network. CC, 2-D image correlation coefficient between the numerically reconstructed image and the original target image. (b) The phase modulation map for the lensless microendoscope of "MST" generated by GS, Core-GS, and CoreNet.

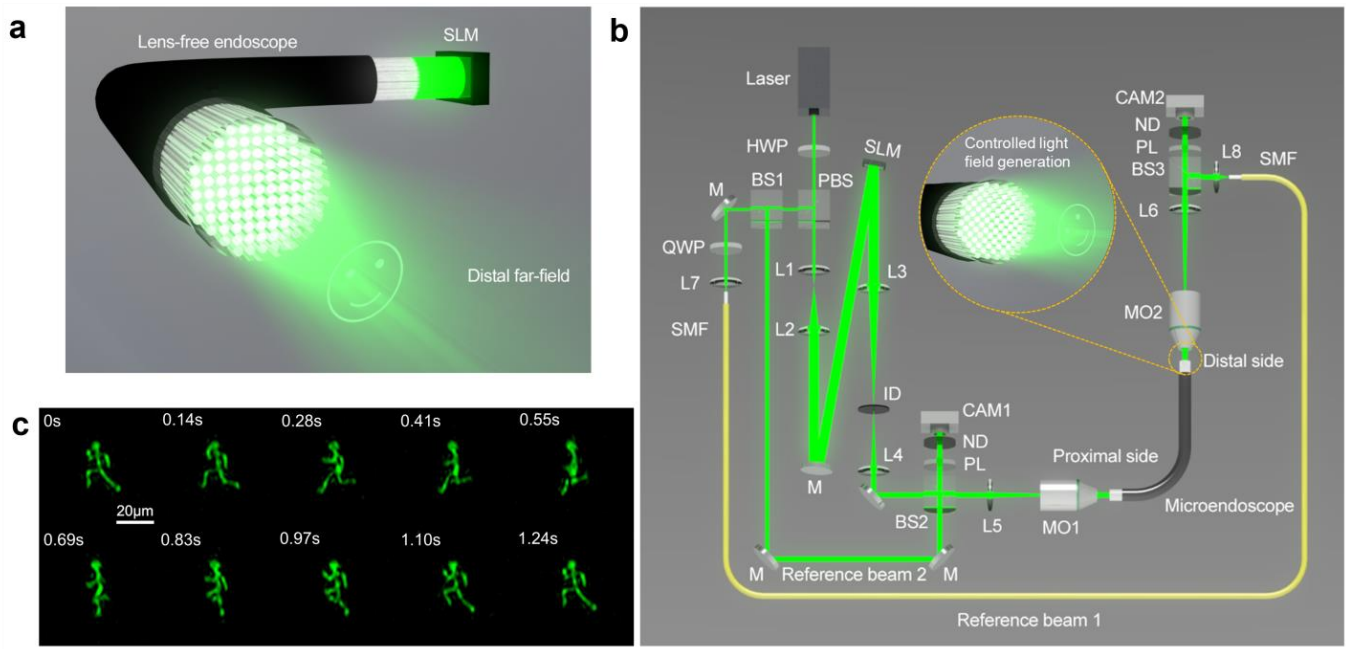

**Fig. 4.** Real-time holographic display through a lensless microendoscope using CoreNet. (a) Schematic demonstrates the principle of controlled light field generation through a 50cm long lensless microendoscope. (b) Sketch of the experimental setup. The Zoomed-in area demonstrates the controlled light field generation at the distal far-field. The phase hologram is generated by CoreNet and projected on the proximal facet of the microendoscope. HWP, half-wave plate; PBS, polarizing beam splitter cube; L1-L8, lenses; BS1-3, beamsplitter cube; QWP, quarter-wave plate; SMF, single-mode fiber; M, mirrors; SLM, spatial light modulator; ID, iris diaphragm; PL, polarizers; ND, neutral density filters; CAM1-2, cameras; MO1-2, microscope objectives. (c) Video-rate tailored light field generation of a running man animation at 700 μm away from the distal fiber facet (see Visualization 1). The tailored holograms are real-time generated by CoreNet and loaded to the phase-only SLM on the fly. The scale bar indicates a length of 20 μm.

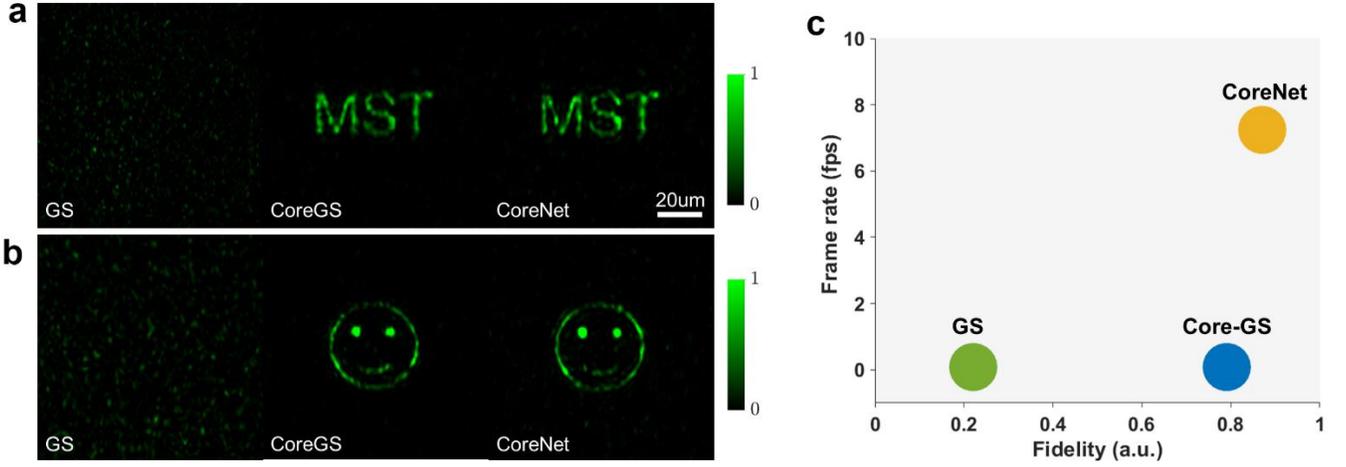

**Fig. 5.** Comparison of GS, Core-GS, and CoreNet. (a) Experimental tailored light field generation of letters "MST" at 700 μm away from the distal fiber facet employing the CGHs generated by GS algorithm, Core-GS algorithm, and CoreNet. The scale bar indicates a length of 20 μm. (b) Experimental tailored light field generation of a smiling face under the same circumstance. (c) Fidelity and computational speed comparison of the three approaches. The fidelity is expressed by the averaged 2-D image correlation coefficients between the reconstructed light field and the target images. The frame rate is the averaged number of generated holograms per second to indicate the computational speed of the algorithm.

### B. Real-time lensless endoscopic light field generation using CoreNet

Employing CoreNet to generate holograms in real-time facilitates the rapid complex holographic display through the lensless microendoscope. The working principle of the lensless microendoscope is shown in Fig. 4a. The holograms generated by CoreNet are loaded to the phase-only SLM in real-time and projected on the proximal fiber facet to generate tailored light fields at the distal far-field. The experimental setup is demonstrated in Fig. 4b, and the detailed description of the setup and calibration process can be found in "Materials and methods". Previously, CGHs needed to be calculated and loaded to the SLM in advance to generate the dynamic light field. Due to the fast computational speed of CoreNet, it is possible to generate the CGHs in real-time for tailored dynamic light field generation through the MCF. As shown in Fig. 4c, a running man animation is reconstructed at the distal far-field of the MCF to demonstrate the rapid hologram generation capability of CoreNet (see Visualization 1). The corresponding modulation holograms are real-time generated by CoreNet and on-the-fly fed to the SLM. To the best of our knowledge, this is for the first time, near-video-rate CGHs generation and holographic display of tailored light field through the miniature lensless microendoscope is demonstrated.

## 3. DISCUSSION

Comparisons of the light field generation in simulation employing CoreNet and the Core-GS are shown in Fig. 3a. Different from the Core-GS [25], CoreNet provides optimal recovery of the target image without any blemishes. It can be noticed in Fig. 3b that the gradient of the phase value in the hologram generated by CoreNet is much smaller than the Core-GS, the smooth transition of the phase leads to homogeneous backgrounds in the reconstructed images, increasing the signal-to-noise ratio.

We use 4 different splits from EMNIST to test the performance of the CoreNet, which is shown in Fig. 6. The EMNIST MNIST and EMNIST Digits dataset provide balanced handwritten digit datasets directly compatible with the original MNIST dataset. The EMNIST Letters dataset merges a balanced set of uppercase and lowercase letters into a single 26-class task. The EMNIST Balanced dataset contains a set of characters with an equal number of samples per class. CoreNet generates accurate CGHs with averaged fidelity over 0.85 for all types of datasets. The performance of CoreNet can be further improved by using other non-pixel-wise losses, such as SSIM loss which could improve the structural similarity or perceptual loss which encourages natural and perceptually pleasing results.

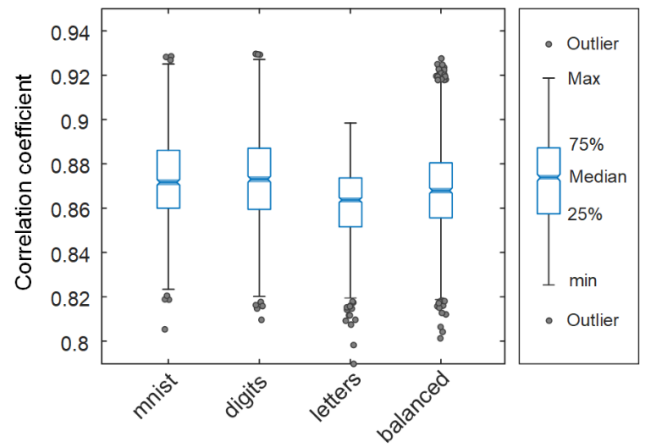

**Fig. 6.** Statistic evaluation of the fidelity for different types of CoreNet output images. The image correlation coefficients are calculated from the numerically reconstructed images and the original target images in over 1000 random tests.

Table 1. Comparison of hologram generation time

| Number of generated holograms | Core-GS | CoreNet |
|---|---|---|
| 1 | 11.1 s | 0.2 s |
| 10 | 115.2 s | 1.4 s |
| 100 | 1170.1 s | 13.8 s |

Comparisons of the light field generation through a 50cm long lensless microendoscope in experiments employing CoreNet and alternate techniques are shown in Fig. 5a. Normal GS algorithm [24] generated holograms lead to a strong distorted light field, even though the phase distortion in the MCF is calibrated in situ. This is mainly due to the random distribution and the limited number of fiber cores, which induces significant spatial aliasing when the modulated light transmits through the MCF. The previously reported Core-GS algorithm solved this problem, providing high-quality complex light field generation through the MCF. However, the iterative process of the Core-GS requires high computational effort. Our novel CoreNet sped up the process by a factor of 82, enabling real-time generation of holograms for rapid holographic display through the microendocope for the first time. The fidelity of the experimental light field generation through the MCF is characterized by employing the correlation coefficients between the experimental captured image and the original target images. The image correlation coefficients, which are calculated between the generated light field in experiments and the target images, for the letter "MST" in Fig. 5a is 0.80 for Core-GS and 0.84 for CoreNet, and for the smiling face in Fig. 5b is 0.67 for Core-GS and 0.69 for CoreNet. Although strong phase distortion is induced by the relatively long length (50 cm) of the lensless microendoscope, the near-perfect in-situ calibration keeps the high fidelity of the light-field generation. Compared to the Core-GS algorithm, CoreNet offers light field generation with higher fidelity with significantly less computation time.

As an unsupervised learning approach, CoreNet provides high-quality phase retrieval for a randomly distributed phased array without labeling. Compared to the Core-GS which is an iterative algorithm, CoreNet significantly reduces the computation time to less than 0.14 s for the generation of one phase hologram, enabling real-time light field generation. The computational speed can be further increased in a better hardware platform. Despite the much shorter calculation time, the generated light field from CoreNet has the highest fidelity in the three approaches (Fig. 5c).

One potential application of CoreNet is in optogenetics. Holographic controlled light enables selective stimulation of target neurons individually with program-controlled shapes of the light field [8,42], enabling precise control of the neuronal networks. The MCF endoscope with a micro-objective (external diameter of 2.6 mm) has been employed for selective photoactivation in a mouse brain with minor invasiveness [32]. However, the low SNR of the generated light field can lead to photoactivation of unwanted neurons, degrading the quality of the selective photoactivation. Holographic stimulation using CoreNet through the MCF can avoid this by generating tailored light fields with high fidelity, enabling in-vivo single-neuron activation. As shown in Fig. 7a, the invasiveness can be minimized to a few hundred microns using the lensless microendoscope for in-vivo selective holographic stimulation. Furthermore, to compensate for the vibration in a behaving mouse, the tailored light field also needs to be controlled adaptively, and CoreNet is the first reported approach that can adaptively control the tailored light field by generating the CGHs in real-time. Closed-loop control of the photoactivation using the lensless microendoscope is thus possible with CoreNet. Hence, the strong capability of rapidly generating high-quality tailored CGHs of CoreNet turns the lensless microendoscope into a powerful optogenetic probe for adaptive and selective photoactivation for in-vivo applications.

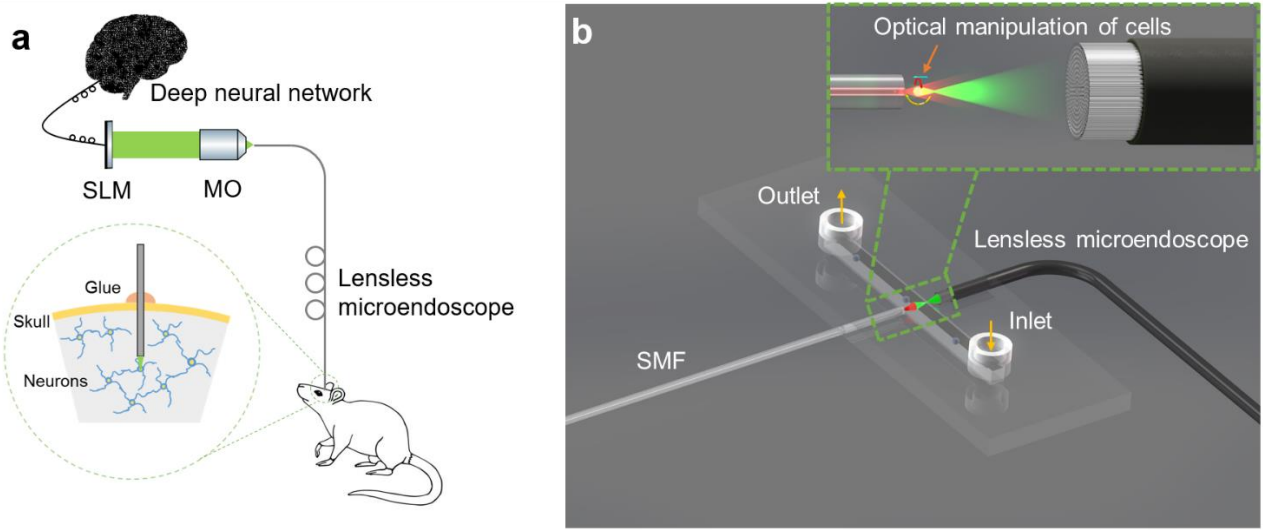

**Fig. 7.** Application of the deep learning enhanced lensless microendoscope. (a) Sketch illustrates employing the lensless microendoscope for holographic photoactivation in mouse brain with minimum invasiveness. The tailored CGHs are generated on the fly by CoreNet. (b) Schematic demonstrates the MCF-based lab-on-a-chip optical manipulation system. CoreNet boosted the generation process of the tailored CGHs for the MCF, providing more degrees of freedom for MCF-based optical manipulation.

Fiber-based optical traps are now an important tool for investigating biological cell mechanics [43–46]. The small size and the high flexibility of optical fibers make them easy to be integrated into miniature lab-on-a-chip devices (Fig. 7b), facilitating high throughput measurements when combined with micro-fluid techniques. We previously reported the first MCF-based dual-beam trap, offering a very high degree of freedom for optical manipulation of biological cells [26]. However, the temporal resolution of optical manipulation is limited by the generation speed of the modulation holograms. Employing CoreNet can boost the generation speed of the CGHs for dynamic light fields. Besides, the refractive index distribution in biological cells is inhomogeneous, leading to instability in optical traps. Adaptive tomographic optical trap [9,31] solved the problem by utilizing tailored trapping beams which fit the refractive index distribution and the shape of cells, but it requires high-speed tailored hologram generation. Hence, it is now possible to implement the adaptive tomographic trap in MCF-based optical traps for real-time closed-loop control. This can significantly increase the stability of the optical trapping and manipulation.

Besides biomedical applications, the MCF is also one of the candidates for the next generation of fiber-optic communication cables. Employing MCFs broaden the bandwidth significantly with high-dimensional communication channels. CoreNet offers the possibility to generate tailored light fields through the MCF in real-time for the first time, paving the way for MCF-based high-dimensional fiber-optic communication [4]. Furthermore, CoreNet can have much wider applications beyond optical engineering. It can be employed for phase retrieval of any kinds of discrete or randomly distributed phased arrays, like phased array radar, ultrasonic phased arrays, opening new perspectives in astronomy, radar technologies, communication technologies, and ultrasonic technologies.

## 4. MATERIALS AND METHODS

### A. Loss function and training details

We use the negative Pearson correlation coefficient (NPCC) as the loss function, which is defined in Eq. 5. The NPCC measures the linear correlation between two images instead of calculating pixel-wise error, which relaxes the constraint on the output image so that the network can converge faster to the optimal solution.

$$L_{\text{NPCC}}(X,Y) = -cc \quad (5)$$

Network training and testing were performed on a workstation with AMD Ryzen 9 3950X CPU and 128 GB of RAM, using NVIDIA RTX A6000 GPU. The network is trained for 5 epochs using the Adam optimizer. The training images are preprocessed to sizes of 512 × 512 pixels and then padded with zeros to 1920 × 1080 pixels.

### B. Experimental setup

The experimental setup is illustrated in Fig. 4b. The diameter of the laser beam emitted from a diode-pumped solid-state laser (Verdi 532 nm, Coherent Inc.) is expanded by a factor of 10 (L1, L2) to fully illuminate the phase-only SLM (PLUTO LCOS SLM, Holoeye Photonics). The CGH displayed on the SLM is combined with a blazed grating, a phase conjugation layer, and a phased array modulation layer. To get rid of the direct reflection from the surface of the SLM, the phase modulation hologram is diffracted to higher orders by the blazed grating and only the first diffraction order can pass through the iris diaphragm in the spatial filter system (L3, ID, L4). The filtered phase modulation hologram is projected on the proximal facet of a 50 cm long MCF (FIGH-350S, Fujikura) through a microscope objective (MO1; 20X Plan Achromat Objective, 0.4 NA, Olympus).

### C. Calibration of the intrinsic phase distortion in MCFs

Before transforming the MCF microendoscope to a phased array, the phase distortion due to the OPD between the fiber cores needs to be compensated employing DOPC [38,47]. In our work, we implement the previously proposed two-stage calibration method [23,48]. The intrinsic OPDs between the fiber cores are measured and compensated in transmission geometry, and the bending induced and temporal phase distortion is further calibrated by the back-reflected guide star in situ. To be more specific, a blazed grating is displayed on the SLM to generate a plane wave illumination at the proximal facet of the MCF, the distorted light field at the distal facet is imaged on the distal camera (CAM2; uEye camera, IDS) through another microscope objective (MO2; 20X Plan Achromat Objective, 0.4 NA, Olympus). A reference beam (reference beam 1 in Fig. 4b) is split from the laser source and coupled into a single-mode fiber. The reference beam from the fiber collimator (L8; collimation package, Thorlabs) is slightly tilted for the digital off-axis holographic geometry [49]. The phase differences of fiber cores are reconstructed from the captured digital hologram on the distal camera [26]. The measured phase is then conjugated and affine transformed into the coordinate system of the SLM to pre-compensate the intrinsic phase distortion.

To calibrate the temporal and bending induced phase distortion in situ, a partial reflector can be mounted on the distal tip, and the further temporal and bending induced phase distortion is measured from the guide star hologram captured on the proximal camera (CAM1; uEye camera, IDS) [48]. The guide star is generated at the distal side by illuminating a single fiber core. The reflected light illuminates the distal facet and the distorted light field on the proximal facet is imaged on the proximal camera interfering with the second reference beam (reference beam 2 in Fig. 4b). Therefore, the phase distortion can be measured from the digital off-axis hologram captured on the proximal camera without distal access. The conjugated phase distortion is then added to the phase conjugation layer of the SLM to further in-situ compensate for the temporal and bending induced phase distortion.

## 5. CONCLUSION

We demonstrated a novel phase retrieval method based on a phase encoder deep neural network (CoreNet) for real-time generation of the tailored light field through the MCF lensless microendoscope. The phase distortion in the MCF is compensated in situ by DOPC, which transforms the MCF to a holographic controlled phased array. Employing CoreNet to generate the tailored holograms for complex wavefront shaping through the MCF provides high fidelity holographic reconstruction at the distal side of the MCF. For the first time, a near video-rate holographic display of dynamic light field is realized through an MCF with on-the-fly generated CGHs. Our work paves the path for high-speed MCF-based applications such as microendoscopic imaging, in-vivo adaptive optical manipulation, optogenetic stimulation, micro-materials processing, and optical communication.

**Funding.** This work is supported by German Research Foundation (DFG) (CZ55/40-1), National Natural Science Foundation of China (NSFC) (62035003, 61775117) and Tsinghua Scholarship for Overseas Graduate Studies (2020023).

**Acknowledgements.** We would like to thank the assistance and valuable discussion from Elias Scharf, Felix Schmieder, Alexander Echeverría Kientzle, and Jakob Dremel.

**Disclosures.** The authors declare no conflict of interest.

**Data availability.** Data underlying the results presented in this paper are available from the corresponding authors upon request.

**Supplementary information.** See Visualization 1 for supporting content.